\documentclass[preprint,aps,prd]{revtex4-1}
\usepackage{graphicx}
\usepackage{textcomp}
\usepackage{amsmath}
\usepackage{amssymb}
\usepackage{bm}
\newcommand{\be}{\begin{equation}}
\newcommand{\ee}{\end{equation}}
\newcommand{\Be}{\begin{eqnarray}}
\newcommand{\Ee}{\end{eqnarray}}
\newcommand{\f}{\frac}
\begin{document}

\title{The Gravitational Perturbation of a Morris-Thorne Wormhole and The Newman-Penrose Formalism}

\author{YuRi Kang}  \author{Sung-Won Kim}
\email[email:]{sungwon@ewha.ac.kr}
\affiliation{Department of Science Education, Ewha Womans University, Seoul 03760, Korea}
\date{today}



\begin{abstract}
The gravitational perturbation of the Morris-Thorne wormhole has been derived by using the Newman-Penrose formalism. We apply Teukolsky equation to the wormhole spacetime, compute the perturbed Weyl scalars, $\Psi_{4}^{(1)}$  and obtain its master equation, decomposed in spin weighted spherical harmonics with spin weight $-2$. For simplicity, we consider the perturbation provoked by a single gaussian pulse of pressureless dust matter.

\end{abstract}

\maketitle

\section{Introduction}
The perturbation theory of spacetime has been one of the most effective tools to understand the physical properties of compact objects in our universe, including black holes and investigate the stability of spacetime around them. A number of studies on the black hole perturbation theory has been conducted to investigate the stability of the black hole spacetime, the quasi-normal modes and the gravitational waves. In the case of Schwarzschild black hole spacetime, the Regge-Wheeler-Zerilli formalism \cite{1, 2} describes the first order metric perturbation and the equations are the single, decoupled equations for the odd and even parity modes. By using the Fourier harmonic expansion, the master equations of the Regge-Wheeler and the Zerilli equation are reduced to the radial ordinary differential equations in the spherically symmetric spacetime. However, in the case of Kerr black hole spacetime, there is no such Regge-Wheeler-Zerilli formalism for the metric perturbation. Instead, the linear perturbation of the Kerr black hole can be described by the Newman-Penrose formalism and the perturbed Weyl scalars, $\psi_0$ and $\psi_4$, obtained by Teukolsky \cite{3}. The metric perturbations of Kerr spacetime are described by coupled partial differential equations so that they should be solved numerically. The other method to compute the perturbation of Kerr spacetime, called CCK formalism, is to construct the metric perturbation from the perturbation of the Weyl scalars, $\psi_0$ and $\psi_4$ \cite{4,5,6}. Using and devloping CCK formalism, the vacuum metric perturbation was computed by Yunes and Gonzalez \cite{7} and the metric perturbation derived by a point particle orbiting around the Schwarzschild black hole was computed by Keidl, Friedman, and Wiseman \cite{8}. Keidl, Shah, and Friedman \cite{9} calculated the metric perturbation, derived by a particle in circular orbit about a Kerr black hole.\\
In this paper, we consider the metric perturbation derived by the Weyl scalar, $\psi_4$ and the metric perturbation of a point particle which is radially falling into a Morris-Thorne wormhole by using the Newman-Penrose formalism. Even though we succeeded to obtain the first-order metric perturbation of the Morris-Thorne Wormhole in the previous study \cite{12} it was limited to the case that the perturbation was only derived by the background matter constituting the wormhole. It is possible that the particles could fall into the wormhole throat or orbit around the throat which may cause the gravitational perturbation, too. The gravitational perturbation equation derived by the linearized theory of gravity was limited because it could not take the exterior matter into account. In order to consider the exterior matter, we decided to use the Newman Penrose Formalism to compute the gravitational perturbation. This is the first step to deal with the metric perturbation of the Morris-Thorne wormhole spacetime via the Newman-Penrose Formalism.\\
This paper is organized as followings. In the first section, we addressed the process we obtained to derive the perturbation equation via the Newman-Penrose Formalism. In Section 2, we applied the previous work in Section 1 to the Morris-Thorne wormhole. In Section 3, we derived the gravitational perturbation in the Morris-Thorne wormhole spacetime, provoked by the exterior matter, the pressureless dust matter, while we summarized our findings and analysis in Section 4.

\section{The Metric Perturbation}

\subsection{The Newman-Penrose Formalism}
The Newman-Perrose formalism is a tetrad formalism which uses the chosen tetrad basis depending on the symmetries of the spacetime and introduces spinor calculus into general relativity. It chose a set of four null vectors, $l^{\mu}$, $k^{\mu}$, $m^{\mu}$, and $m^{*\mu}$ to a basis at each point of the spacetime. The first two null vectors, $l^{\mu}$ and $k^{\mu}$, are real quantities along the light cone, oriented towards the future. In the asymptotic region of a hypersurface of constant time, $l^{\mu}$ points radially outward and $k^{\mu}$ points radially inward. The last two null vectors, $m^{\mu}$ and $m^{*\mu}$, are usually defined as a complex one and its complex conjugate. The tetrad of null vectors are required to satisfy the orthogonality conditions
\Be
& l \cdot m = l \cdot m^{*} = k \cdot m = k \cdot m^{*} = 0,\\
& l \cdot l = k \cdot k = m \cdot m = m^{*} \cdot m^{*} = 0,\\
& l \cdot k = -m \cdot m^{*} = 1.
\Ee

The metric tensor $g_{\mu \nu}$ of the four dimensional spacetime in terms of null vectors can be written as
\be
g_{\mu \nu} = l_{\mu} k_{\nu} +k_{\mu} l_{\nu} - m_{\mu} m_{\nu}^{*} -m_{\mu}^{*} m_{\nu} .
\ee
Also the metric can be expressed as
\be
g_{\mu\nu} = \eta^{(a)(b)} e_{(a)({\mu}} e_{(b)\nu)}.
\ee
where the flat metric $\eta_{(a)(b)}$ is a constant symmetric matrix, given by
\be
\eta_{(a)(b)} =\left(
                    \begin{array}{cccc}
                    0 & -1 & 0 & 0 \\ -1 & 0 & 0 & 0 \\ 0 & 0 & 0 & 1 \\ 0 & 0 & 1 & 0 \end{array}
                    \right)
\ee
We define the directional operators
\be
D=l^{\mu} \partial_{\mu} ,~~~~~
\bigtriangleup=k^{\mu} \partial_{\mu},~~~~~
\delta=m^{\mu} \partial_{\mu}.
\ee
The Ricci rotation-coefficients, which is also called as the spinor coefficients are given as
\Be
& \kappa_{s} = \gamma_{311} = m^{\mu} l_{\mu ; \nu} l^{\nu};~~~~~&  \tau{s} = \gamma_{312} = m^{\mu} l_{\mu ; \nu} k^{\nu}; \nonumber \\
& \sigma_{s} = \gamma_{313}=m^{\mu} l_{\mu ; \nu} m^{\nu};~~~~~&  \rho{s} = \gamma_{314} = m^{\mu} l_{\mu ; \nu} m^{* \nu}; \nonumber \\
&  \pi_{s} = \gamma_{241} = k^{\mu} m_{\mu ; \nu}^{*} l^{\nu}; ~~~~~& \nu_{s} = \gamma_{242}=k^{\mu} m_{\mu ; \nu}^{*} k^{\nu}; \nonumber \\
& \mu_{s} = \gamma_{243} = k^{\mu} m_{\mu ; \nu}^{*} m^{\nu};~~~~~& \lambda_{s} = \gamma_{244}=k^{\mu} m_{\mu ; \nu}^{*} m^{*\nu}; \nonumber
\Ee
\Be
\epsilon_{s} &=& \f{1}{2}\left(\gamma_{211}+\gamma_{341} \right) = \f{1}{2} \left(k^{\mu} l_{\mu ; \nu} + m^{\mu} m_{\mu ; \nu}^{*} \right) l^{\nu} ; \nonumber \\
\gamma_{s} &=& \f{1}{2}\left(\gamma_{212}+\gamma_{342} \right) = \f{1}{2} \left(k^{\mu} l_{\mu ; \nu} + m^{\mu} m_{\mu ; \nu}^{*} \right) k^{\nu} ; \nonumber \\
\beta_{s} &=& \f{1}{2}\left(\gamma_{213}+\gamma_{343} \right) = \f{1}{2} \left(k^{\mu} l_{\mu ; \nu} + m^{\mu} m_{\mu ; \nu}^{*} \right) m^{\nu} ; \nonumber \\
\alpha_{s} &=& \f{1}{2}\left(\gamma_{214}+\gamma_{344} \right) = \f{1}{2} \left(k^{\mu} l_{\mu ; \nu} + m^{\mu} m_{\mu ; \nu}^{*} \right) m^{* \nu} ,
\Ee
where we add a subindex, ${}_{s}$ to avoid confusion with other symbols.
We want to put an emphasis on the important property with regard to the $l \leftrightarrow k$ exchange transformation/ When we exchange the two real null vectors, the spin coefficients become transformed as
\Be
& \kappa \leftrightarrow -\nu^{*}, ~~~~~   \rho \leftrightarrow -\mu^{*},~~~~~  \sigma \leftrightarrow -\lambda^{*} ,\nonumber \\
& \alpha \leftrightarrow -\beta^{*}, ~~~~~ \epsilon \leftrightarrow -\gamma^{*} , ~~~~~ \pi \leftrightarrow -\tau^{*} .
\Ee
The spin coefficients are the keys to explain the physical properties of the tetrad vectors. Using these exchange properties, we can restrict out attention only to the physical meaning of the spin coefficients which are related to the radially outward null vector, $\textbf{l}$. \\
In Newman-Penrose formalism, the Weyl scalar is the most relevant quantity besides the spin coefficients. The Weyl scalars contain all the information about the spacetime and the choice of the coordinate system does not affect them. They only depend on the choice of the null tetrad because they are scalar quantities.\\
The Weyl tensor is used to compute the Weyl scalar. The Weyl tensor is the trace-free part of the Riemann tensors,
\be
C_{\mu \nu \lambda \delta} = R_{\mu \nu \lambda \delta} -\f{1}{2} \left
( g_{\mu \lambda} R_{\nu \delta} +g_{\nu \delta} R_{\mu \lambda} -g_{\nu \lambda} R_{\mu \delta} \right) + \f{1}{6} \left( g_{\mu \lambda} g_{\nu \delta} - g_{\nu \lambda } g_{\mu \delta} \right) R.
\ee
The Weyl tensor possesses the same symmetries as the curvature tensor. We project the Weyl tensor on to the tetrad frame and the tetrad components are
\Be
R_{(a)(b)(c)(d)} &=& C_{(a)(b)(c)(d)}\nonumber \\
&& -\f{1}{2} \left(\eta_{(a)(c)} R_{(b)(d)} -\eta_{(b)(c)} R_{(a)(d)} -\eta_{(a)(d)} R_{(b)(c)} +\eta_{(b)(d)} R_{(a)(c)} \right) \nonumber \\
&& +\f{1}{6} \left(\eta_{(a)(c)} \eta_{(b)(d)} -\eta_{(a)(d)} \eta_{(b)(c)} \right)R,
\Ee
where $R_{(a)(c)}$ denotes the tetrad components of the Ricci tensor and $R$ denotes that of the scalar curvature.\\
In the Newman-Penrose formalism, the five complex scalars are used to represent the ten independent components of the Weyl tensor.
\Be
\Psi_{0} &=& C_{(1)(3)(1)(3)} = - C_{\mu \nu \lambda \delta} l^{\mu} m^{\nu} l^{\lambda} m^{\delta} , \nonumber \\
\Psi_{1} &=& - C_{(1)(2)(1)(3)} = - C_{\mu \nu \lambda \delta} l^{\mu} k^{\nu} l^{\lambda} m^{\delta} , \nonumber \\
\Psi_{2} &=& - C_{(1)(3)(4)(2)} = - C_{\mu \nu \lambda \delta} l^{\mu} m^{\nu} m^{* \lambda} k^{\delta} , \nonumber \\
\Psi_{3} &=& - C_{(1)(2)(4)(2)} = - C_{\mu \nu \lambda \delta} l^{\mu} k^{\nu} m^{* \lambda} k^{\delta} , \nonumber \\
\Psi_{4} &=& - C_{(2)(4)(2)(4)} = - C_{\mu \nu \lambda \delta} k^{\mu} m^{* \nu} k^{\lambda} m^{* \delta}.
\Ee
The Weyl scalars are not definite quantities and their values can be changed via the tetrad transformation.\\
We can subject the tetrad frame to a Lorentz transformation at some point and extend it continuously throughout the entire spacetime. Therefore, the tetrad has six degrees of freedom such as \\
(1) rotations about the null vector $l$
\Be
l &\rightarrow& l, \nonumber \\
m &\rightarrow& m + a l, \nonumber \\
k &\rightarrow& k + a^{*} m +a m^{*} +a a^{*} l,
\Ee
(2) rotations about the null vector $k$
\Be
k &\rightarrow& k, \nonumber \\
m &\rightarrow& m + b k, \nonumber \\
l &\rightarrow& l + b^{*} m +b m^{*} +b b^{*} l,
\Ee
(3) rotations about $m$, rotating $m$ and $m^{*}$ by an angle $\theta$ in the $m-m^{*}$ plane while keeping $l$ and $k$ fixed
\be
m \rightarrow e^{\imath \theta} m,
\ee
where $a$ and $b$ are two complex functions and $\theta$ is real function on the manifold.\\
This freedom introduces the concept of spin weight, $s$ and a quantity that goes through rotation (3) is said to have spin weight $s$.
\be
\xi \rightarrow e^{\imath s \theta} \xi
\ee

\subsection{Perturbation Equations}
In 1973, Teukolsky derived the linearized perturbation equations, applicable to any type D spacetime in Petrov classification by using the Newman-Penrose formalism. Teukolsky specified the perturbed geometry by
\Be
l &=& l^{A} + l^{B}, \nonumber \\
k &=& k^{A} + k^{B}.
\Ee
The label $A$ represents the background spacetime and the label $B$ indicates the perturbed spacetime. We decided to keep $B$ terms only to the first order.\\
With a suitable tetrad basis,
\Be
\Psi_{0}^{A} = \Psi_{1}^{A} = \Psi_{3}^{A} = \Psi_{4}^{A} =0, \nonumber \\
\kappa^{A} = \sigma^{A} = \nu^{A} = \lambda^{A} =0.
\Ee
Teukolsky combined the non-vanishing Bianchi identities and Ricci identities for the first-order perturbation of the radiation scalars $\Psi_{0}^{B}$ and $\Psi_{4}^{B}$ and obtained the following two decoupled equations:
\Be
\left[ \left( D - 3\epsilon + \epsilon^{*} -4\rho - \rho^{*} \right) \left(\Delta - 4\gamma + \mu \right) \right.  \nonumber \\
\left. - \left(\delta+ \pi^{*} - \alpha^{*} - 3 \beta - 3 \tau \right)\left(\delta^{*} +\pi -4\alpha \right)-3 \Psi_{2}^{A} \right] \Psi_{0}^{B} = 4 \pi T_{0} , \\
\left[ \left( \Delta + 3 \gamma - \gamma^{*} + 4 \mu + \mu^{*} \right) \left( D +4 \epsilon - \rho \right) \right. \nonumber \\
\left. - \left(\delta^{*} - \tau^{*} + \beta^{*} + 3 \alpha + 4 \pi \right) \left(\delta - \tau +3 \beta \right) - 3 \Psi_{2}^{A} \right] \Psi_{4}^{B} = 4 \pi T_{4},
\Ee

where $D$, $\Delta$, $\delta$ and $\delta^{*}$ are the directional derivatives defined as
\Be
D &=& l^{\mu} \nabla_{\mu} , ~~~~~~~~~~\Delta = k^{\mu} \nabla_{\mu}, \nonumber \\
\delta &=& m^{\mu} \nabla_{\mu}, ~~~~~~~~~~\delta^{*} = m^{* \mu} \nabla_{\mu},
\Ee
and
\Be
T_{0} &=& \left(\delta + \pi^{*} -\alpha^{*} -3 \beta - 4\pi \right) \left[ \left( D-2 \epsilon - 2 \rho^{*} \right) T_{lm}^{B} - \left( \delta + \pi^{*} -2 \alpha ^{*} - 2 \beta \right) T_{ll}^{B} \right] \nonumber \\
&&+\left( D- 3 \epsilon + \epsilon^{*} -4 \rho - \rho^{*} \right)\left[\left( \delta + 2 \pi^{*} - 2 \beta \right) T_{lm}^{B} - \left(D- 2 \epsilon + 2 \epsilon^{*}  - \rho^{*} \right) T_{mm}^{B} \right], \nonumber \\
T_{4} &=& \left(\Delta + 3 \gamma - \gamma^{*} + 4 \mu + \mu^{*} \right) \left[ \left( \delta^{*} - 2 \tau^{*} +2 \alpha \right) T_{km^{*}} - \left( \Delta + 2\gamma - 2\gamma^{*} +\mu^{*} \right) T_{m^{*} m^{*}} \right] \nonumber \\
&&+\left( \delta^{*} - \tau^{*} + \beta^{*} + 3 \alpha + 4\pi \right) \left[ \left( \Delta + 2 \gamma + 2 \mu^{*} \right) T_{km^{*}} - \left( \delta^{*} - \tau^{*} + 2 \beta^{*} + 2 \alpha \right) T_{kk} \right]. \nonumber
\Ee
We computed the master equation of Teukolsky equations for the perturbed Weyl scalar $\Psi_{4}$, including the source terms
\Be
\left[ \left( \Delta + \eta \left( 4\mu_{s} + \mu_{s}^{*} + 3 \gamma_{s} - \gamma_{s}^{*} \right) \right) \left( D- \eta \left( \rho_{s} - 4 \epsilon_{s} \right) \right) \right. \nonumber \\
\left. -\left( \delta^{*} + \eta \left( 3 \alpha_{s} + \beta_{s}^{*} + 4 \pi_{s} -\tau_{s}^{*} \right) \right) \left( \delta + \eta \left( 4 \beta_{s} -\tau_{s} \right) \right) -3 \eta \Psi_{2} \right] \Psi_{4}^{1} = \eta \f{K}{2} T_{4},
\Ee
where $K=8\pi$ in geometrized units and $\eta = -1$ by using the signature $(-, +, +, +)$.\\
This master equation describes the linear order dynamics of a gravitationally perturbed spacetime.\\
The source terms, $T_{4}$, are
\be
T_{4} = {\hat{T}}^{kk} T_{kk} + {\hat{T}}^{km^{*}} T_{km^{*}} + {\hat{T}}^{m^{*} m^{*}} T_{m^{*} m^{*}} ,
\ee
where ${\hat{T}}^{ab}$ is the operators acting on the projections of the perturbed source term $T_{\mu \nu}$ on the null tetrad,
\Be
{\hat{T}}^{kk} &=& -\left( \delta^{*} + \eta \left( 3 \alpha_{s} + \beta_{s}^{*} + 4\pi_{s} - \tau_{s}^{*} \right) \right) \left( \delta^{*} + \eta \left( 2 \alpha_{s} + 2 \beta_{s}^{*} - \tau_{s}^{*} \right) \right) , \nonumber \\
{\hat{T}}^{k m^{*}} &=& \left( \Delta + \eta \left( 4 \mu_{s} + \mu_{s}^{*} + 3 \gamma_{s} - \gamma_{s}^{*} \right) \right) \left( \delta^{*} + 2 \eta \left( \alpha_{s} - \tau_{s}^{*} \right) \right) \nonumber \\
&& + \left( \Delta + 2 \eta \left(\mu_{s}^{*}  + \gamma_{s}\right) \right) \left( \delta^{*} + \eta \left( 3 \alpha_{s} + \beta_{s}^{*} + 4 \pi_{s} - \tau_{s}^{*} \right) \right), \nonumber \\
{\hat{T}}^{m^{*} m^{*} } &=& -\left( \Delta + \eta \left( 4\mu_{s} + \mu_{s}^{*} +3 \gamma_{s} - \gamma_{s}^{*} \right) \right) \left( \Delta + \eta \left( \mu_{s}^{*} + 2 \gamma_{s} - 2 \gamma_{s}^{*} \right) \right).
\Ee

\section{The Morris-Thorne wormhole and the Newman-Penrose Formalism}
In Schwarzschild coordinate $(t, r, \theta \phi)$, the Morris-Thorne wormhole metric is given by
\be
ds^{2} = -e^{2 \Phi_{\pm} (r)} dt^{2} + \f{dr^2}{1-\f{b_{\pm}(r)}{r}} + r^{2} \left( d\theta^{2} + sin^{2} \theta d \phi^{2} \right),
\ee
where $Psi_{r}$ and $b(r)$ are the red-shift function and the shape function, respectively. \\
We first consider the null vectors, representing the radial null-geodesics
\Be
l^{\mu} &=& \left(l^{t} , l^{r} , l^{\theta}, l^{\phi} \right) = \f{1}{2} \left( e^{-2\Phi} , e^{-\Phi} \left(1-\f{b}{r} \right)^{\f{1}{2}}, 0, 0 \right), \nonumber \\
k^{\mu} &=& \left( k^{t}, k^{r}, k^{\theta}, k^{\phi} \right) = \left(1, -e^{\Phi} \left( 1-\f{b}{r} \right)^{\f{1}{2}} ,0,0 \right),
\Ee
which satisfy the required orthogonality condition $l\cdot k =\eta = -1$. \\
We now consider the complex null vector
\Be
m^{\mu} &=& \f{1}{\sqrt{2} r} \left(0, 0, 1, \imath \csc{\theta} \right),\nonumber \\
m^{*\mu} &=& \f{1}{\sqrt{2} r} \left( 0, 0, 1, -\imath \csc{\theta} \right),
\Ee
which is orthogonal to $l$ and $k$ and satisfy the normalization $m \cdot m^{*} = -\eta = 1$. \\
The directional derivatives are
\Be
D &=& l^{\mu} \nabla_{\mu} = \f{1}{2} e^{-2\Phi} \left[ \f{\partial}{\partial t} + e^{\Phi} \left( 1 - \f{b}{r}\right)^{\f{1}{2}} \f{\partial}{\partial r} \right] , \nonumber \\
\Delta &=& k^{\mu} \nabla_{\mu} = \f{\partial}{\partial t } -e^{\Phi} \left( 1- \f{b}{r} \right)^{\f{1}{2}} \f{\partial}{\partial r}, \nonumber \\
\delta &=& m^{\mu} \nabla_{\mu} = \f{1}{\sqrt{2} r} \left( \f{\partial}{\partial \theta} +\f{\imath}{\sin{\theta}} \f{\partial}{\partial \phi} \right), \nonumber \\
\delta^{*} &=& m^{*\mu} \nabla_{\mu} = \f{1}{\sqrt{2} r} \left( \f{\partial}{\partial \theta} - \f{\imath}{\sin{\theta}} \f{\partial}{\partial \phi} \right).
\Ee
The non-vanishing spin coefficients are
\Be
\rho_{s} &=& \f{(r-b)}{2 r^{2}} e^{-\Phi} \left( 1- \f{b}{r} \right)^{-\f{1}{2}},
 ~~~~~~ \mu_{s} =\f{e^{\Phi}}{r} \left( 1-\f{b}{r} \right)^{\f{1}{2}},
 ~~~~~~ \gamma_{s} = -\Phi ' e^{\Phi} \left( 1 - \f{b}{r} \right)^{\f{1}{2}}, \nonumber \\
 \beta_{s} &=& -\f{1}{2 \sqrt{2} r} \f{\cos{\theta}}{\sin{\theta}},
 ~~~~~~~~~~~~\alpha_{s} = \f{1}{2 \sqrt{2} r} \f{\cos{\theta}}{\sin{\theta}},\nonumber \\
 \kappa_{s} &=& \tau_{s} =\alpha_{s} =\pi_{s} =\nu_{s} =\lambda_{s} =\epsilon_{s} = 0.
 \Ee
 From the fact that the spin coefficients $\kappa$, $\sigma$, $\lambda$, and $\nu$ vanish, we concluded that the Morris-Thorne wormhole spacetime is a type-D spacetime on the basis of the Goldberg-Sachs theorem \cite{10}. According to the theorem, we noticed that the Weyl scalars $\Psi_{0}$, $\Psi_{1}$, $\Psi_{3}$ and $\Psi_{4}$ vanish and that the non-vanishing scalar is $\Psi_{2}$ in the chosen basis. \\
 The master equation of perturbation is
 \be
 \left[\Box_{tr}^{\Psi} +\Box_{\theta \phi} \right] \Psi_{4}^{(1)} = K r^{2} T_{4},
 \ee
 where
 \Be
 \Box_{tr}^{\Psi} &=& -r^{2} e^{-2\Phi} \f{\partial^{2}}{\partial t^{2} } + r (r-b) \f{\partial^{2}}{\partial r^{2}} \nonumber \\
 && - 4 e^{-\Phi} (r^{2} - b r)^{1/2} (r \Phi ' - 1) \f{\partial}{\partial t} - \left[ \f{( b' r-b)}{2} +3( \Phi ' r-2)( r-b ) \right] \f{\partial}{\partial r} \nonumber \\
 && - 3 (3 \Phi' r -1) \left( 1- \f{b}{r} \right) +( 1- b') + \f{7( b'r-b)}{2r} ,\nonumber \\
 \Box_{\theta \phi} &=& \f{\partial^{2}}{\partial \theta^{2}} +\f{1}{\sin^{2}{\theta}} \f{\partial^{2}}{\partial \phi^{2}} + \f{\cos{\theta}}{\sin{\theta}} \f{\partial}{\partial \theta} - 4 \imath \f{\cos{\theta}}{\sin^{2} {\theta}} \f{\partial}{\partial \phi} -\f{2}{\sin^{2} {\theta}} ( 1+ \cos^{2} {\theta} ).
 \Ee
On the right-hand side, the source term $T_{4}$ take the form with the explicit form of the operators
\Be
{\hat{T}}^{kk} &=& -\f{1}{2 r^{2}} {\bar{\partial}}_{-1} {\bar{\partial}}_{0}, \nonumber \\
{\hat{T}}^{km^{*}} &=& - \f{1}{\sqrt{2}} \left[ \f{2}{r} \left[ \f{\partial}{\partial t} -e^{\Phi} \left( 1-\f{b}{r} \right)^{\f{1}{2}} \f{\partial}{\partial r} \right] - \f{e^{\Phi}}{r} \left( 1- \f{b}{r} \right)^{\f{1}{2}} \left( \f{5}{r} -4\Phi' \right) \right] {\bar{\partial}}_{-1}, \nonumber \\
{\hat{T}}^{m^{*} m^{*}} &=& - \left[ \f{\partial^{2}}{\partial t^{2}} -2 e^{\Phi} \left( 1- \f{b}{r} \right)^{\f{1}{2}} \f{\partial^{2}}{\partial t \partial r} + e^{2 \Phi}  \left( 1-\f{b}{r} \right) \f{\partial^{2}}{\partial r^{2}} \right. \nonumber \\
&& -2 e^{\Phi} \left( 1-\f{b}{r} \right)^{\f{1}{2}} \left( \f{3}{r} - \Phi ' \right) \f{\partial}{\partial t} + e^{2 \Phi} \left[ \left( 1- \f{b}{r} \right) \left( \f{6}{r} - \Phi ' \right) - \f{(b'r-b)}{2r^{2}} \right] \f{\partial}{\partial r} \nonumber \\
&& \left. +e^{2 \Phi } \left[ \left( 1- \f{b}{r} \right) \left( \f{4}{r^{2}} - \f{\Phi'}{r} \right) - \f{(b'r-b)}{2 r^{3}} \right] \right] .
\Ee
We used the operators $\partial_{s}$ and ${\bar{\partial}}_s$ defined as
\Be
\partial_{s} &=& -\left( \f{\partial}{\partial \theta} + \f{\imath}{\sin{\theta}} \f{\partial}{\partial \phi} - s \cot{\theta} \right)\equiv \partial_{0} + s \cot{\theta}, \nonumber \\
{\bar{\partial}}_{s} &=& -\left( \f{\partial}{\partial \theta} -\f{\imath}{\sin{\theta}} \f{\partial}{\partial \phi} + s \cot{\theta} \right) \equiv {\bar{\partial}}_{0} -s \cot{\theta} ,
\Ee
in which the subindex $s$ indicates that these operators act on quantities of spin weights. When the operators acts on the spin weighted spherical harmonics $ Y_{s}^{l,m} (\theta, \phi)$, they raise and lower the spin weight of the harmonic so that they are called as the spin raising and lowering operators, respectively.
\Be
\partial_{s} Y_{s}^{l,m} &=& \sqrt{(l-s)(l+s+1)} Y_{s+1}^{l,m} ,\nonumber \\
{\bar{\partial}}_{s} Y_{s}^{l,m} &=& - \sqrt{(l+s)(l-s+1)} Y_{s-1}^{l,m}.
\Ee

We consider the case of $\Phi(r)=0$ and $b(r)=\f{b_{0}^{2}}{r} $ ($b_{0}$ is a constant) for simplicity and rewrite the final master equation for the perturbed scalar $\Psi_{4}$
\be
\left[\Box_{tr}^{\Psi} +\Box_{\theta \phi} \right] \Psi_{4}^{(1)} = K r^{2} T_{4}, \nonumber \\
\ee
\Be
\Box_{tr}^{\Psi} &=& -r^{2} \f{\partial^{2}}{\partial t^{2}} + (r^{2} -b_{0}^{2} ) \f{\partial^{2}}{\partial r^{2}} + 4 (r^{2} - b_{0}^{2} )^{\f{1}{2}} \f{\partial}{\partial t} + r \left( 6-5 \f{b_{0}^{2}}{r^{2}} \right) \f{\partial}{\partial r} +\left( 4- 9 \f{b_{0}^{2}}{r^{2}} \right), \nonumber \\
\Box_{\theta \phi} &=& \f{\partial^{2}}{\partial \theta^{2}} +\f{1}{\sin^{2}{\theta}} \f{\partial^{2}}{\partial \phi^{2}} +\f{\cos{\theta}}{\sin{\theta}} \f{\partial}{\partial \theta} -4\imath \f{\cos{\theta}}{\sin^{2}{\theta}} \f{\partial}{\partial \phi} -\f{2}{\sin^{2}{\theta}} (1+\cos^{2}{\theta}).
\Ee
We also rewrote the source term, $T_{4}$,
\Be
{\hat{T}}^{kk} &=& -\f{1}{2 r^{2}} {\bar{\partial}}_{-1} {\bar{\partial}}_{0}, \nonumber \\
{\hat{T}}^{km^{*}} &=& -\f{1}{\sqrt{2}} \left[ \f{2}{r} \left[ \f{\partial}{\partial t} - \left( 1- \f{b_{0}^{2}}{r^{2}} \right)^{\f{1}{2}} \f{\partial}{\partial r} \right] - \f{5}{r^{2}} \left( 1- \f{b_{0}}{r^{2}} \right)^{\f{1}{2}} \right] {\bar{\partial}}_{-1}, \nonumber \\
{\hat{T}}^{m^{*} m^{*}} &=& -\left[ \f{\partial^{2}}{\partial t^{2}} +\left( 1- \f{b_{0}^{2}}{r^{2}} \right) \f{\partial^{2}}{\partial r^{2}} -2 \left( 1- \f{b_{0}^{2}}{r^{2}} \right)^{\f{1}{2}} \f{\partial^{2}}{\partial t \partial r} \right. \nonumber \\
&& \left. ~~~~~-\f{6}{r} \left( 1-\f{b_{0}^{2}}{r^{2}} \right) \f{\partial}{\partial t} +\f{1}{r} \left( 6-7 \f{b_{0}^{2}}{r^{2}} \right) \f{\partial}{\partial r} +\f{1}{r^{2}} \left(4-3 \f{b_{0}^{2}}{r^{2}} \right) \right] .
\Ee
We chose to set the function $ r\Psi_{4}^{(1)} = \Phi_{4}^{(1)}$ so that the master equation becomes
\be
\left[ \Box_{tr}^{\Phi_{4}} +\Box_{\theta \phi} \right] \Phi_{4} = Kr^{3} T_{4} \nonumber
\ee
where
\Be
\Box_{tr}^{\Phi_{4}} &=& -r^{2} \f{\partial^{2}}{\partial t^{2}} + (r^{2} -b_{0}^{2} ) \f{\partial^{2}}{\partial r^{2}} +4(r^{2} -b_{0}^{2} )^{\f{1}{2}} \f{\partial}{\partial t} + r \left( 4- \f{3b_{0}^{2}}{r^{2}} \right) \f{\partial}{\partial r} - 6 \f{b_{0}^{2}}{r^{2}}, \nonumber \\
\Box_{\theta \phi} &=& {\bar{\partial}}_{-1} \partial_{-2}.
\Ee

Because the spin operators act on the spin weighted spherical harmonic functions, we could decompose the Weyl scalar $\Psi_{4}^{(1)}$ into the temporal-radial part and the angular part by using a spherical harmonics with spin weight $-2$
\be
\Phi_{4}^{(1)} =\sum_{lm} {R_{l,m} (t,r) Y_{-2}^{lm} (\theta, \phi)}.
\ee
Therefore, we finally obtained the master equation for the gravitational perturbation of a Morris-Thorne wormhole in the case of $\Phi(r)=0$ and $b(r)=\f{b_{0}^{2}}{r} $
\be
\Box_{tr}^{\Phi_{4}^{(1)}} R_{lm} -(l-1)(l+2) R_{lm} = K r^{3} T_{4}.
\ee

\subsection{The Simplest Morris-Thorne wormhole, $\Phi(r)=0$ and $b(r)=\f{b_{0}^{2}}{r} $ }
In the Morris-Thorne wormhole spacetime, the throat of the wormhole is located at $r(r_{0} )=b_{0}$. The master equation is regular in the all region, except for $r=0$. This implies that we might be able to see what happens inside the throat $(0<r<r_{0} =b_{0})$ depending on the source term $T_{4}$ on the right-hand side of the equation. When the source term $T_{4}$ has singularities at the throat of the wormhole, we should adapt the general coordinate of the traversable wormhole back. With the proper radial distance,
\be
r^{*}(r)=\pm \int_{r_{0}}^{r}{\f{dr}{\sqrt{1-\f{b_{\pm} (r)}{r}}}},
\ee
the perturbed equation became surprisingly simple
\be
\Box_{tr^{*}}^{\Phi_{4}^{(1)}} R_{lm} - (l-1)(l+2) R_{lm} =K r(r^{*})^{3} T_{4},
\ee
where
\be
\Box_{tr^{*}}^{\Phi_{4}^{(1)}} = -(r^{*2} +b_{0}^{2} ) \f{\partial^{2}}{\partial t^{2}} + \left( r^{*2} +b_{0}^{2} \right) \f{\partial^{2}}{\partial r^{*2}} +4 r^{*} \left( \f{\partial}{\partial t} + \f{\partial}{\partial r^{*}} \right) - \f{6 b_{0}^{2}}{(r^{*2} +b_{0}^{2})} .
\ee
The left-hand side of the perturbed equation is regular in all the region even at the throat of throat of the wormhole, $r^{*} = 0$. In addition, the equation is second-order linear partial differential equation so that we can solve it numerically.\\

We considered the case in which there is not external disturbance, which is $T_{4}=0$. The right-hand side of the master perturbed equation becomes zero so that it becomes the homogeneous second-order partial differential equation. Even though it seems to be simple, one cannot solve it analytically because of the first order time and radial derivative terms exist. \\
We assumed that the perturbation is harmonic in order to get rid of the time dependence by using a standard Fourier decomposition, which is proportional to $e^{-\imath \omega t}$.
\be
R_{lm} (t, r^{*} )= \f{1}{2 \pi} \int_{-\infty}^{+\infty} {e^{-\imath \omega t} \xi_{lm} (\omega' , r^{*}) d\omega}.
\ee
The equation became
\be
\f{\partial^2 \xi_{lm}^{\omega}}{dr^{*2}} +\f{4 r^{*}}{(r^{*2} +b_{0}^{2} )} \f{\partial \xi_{lm}^{\omega}}{\partial r^{*}} +\left( \omega^{2} - \f{4 \imath r^{*} \omega +(l-1)(l+2)}{(r^{*2} +b_{0}^{2})} - \f{6 b_{0}^{2}}{(r^{*2} +b_{0}^{2} )^{2}} \right) \xi_{lm}^{\omega} =0.
\ee
The equation can be rewritten in the form of general Sturm-Liouville eigenvalue problem
\be
\f{1}{(r^{*2} + b_{0}^{2} )} \f{d}{d r^{*}} \left[ (r^{*2} +b_{0}^{2} )^{2} \f{d \xi_{lm}^{\omega}}{dr^{*}} \right] = - \left( - \f{\omega^{2} (r^{*2} + b_{0}^{2} )^{2} -6 b_{0}^{2}}{(r^{*2} + b_{0}^{2} ) } + 4 \imath r^{*} \omega +(l-1)(l+2) \right) \xi_{lm}^{\omega}.
\ee
We simplified the equation by defining
\be
\xi_{lm}^{\omega} \equiv \f{\chi (r^{*})}{(r^{*2} +b_{0}^{2} )} ,
~~~~~~~~~~i.e. ~~R_{lm} (t, r^{*} ) = e^{-\imath \omega t} (r^{*2} +b_{0}^{2} )^{-1} \chi_{lm}^{\omega},
\ee
and the perturbed equation became
\be
\f{d^{2} \chi_{lm}^{\omega} }{dr^{*2}} + \left(\omega^{2} -\f{4 \imath \omega r^{*} +(l-1)(l+2)}{r^{*2} +b_{0}^{2}} -\f{6 b_{0}^{2} }{(r^{*2} +b_{0}^{2} )^{2}} \right) \chi_{lm}^{\omega} = 0.
\ee
In a distant region far from the throat of the wormhole $(r^{*} \rightarrow \infty )$, the equation became
\be
\f{d^{2} \chi_{lm}^{\omega}}{dr^{*2}} +\left( \omega^{2} -\f{4 \imath \omega}{r^{*}} \right) \chi_{lm}^{\omega} \approx 0.
\ee
The asymptotic solutions are $X_{lm}^{\omega} \sim r^{* \mp 2} e^{\mp \imath \omega r^{*}}$. The minus sign in the exponential term indicates the ingoing waves while the plus sign in the exponential term indicates the outgoing waves. We can obtain the order of the gravitational perturbation $\Psi_{4}^{(1)}$ in a distant region
\Be
\textrm{(outgoing waves)}~~~~~~~~~~ \Psi_{4}^{(1)} & \sim & \f{e^{\imath \omega r^{*}}}{r^{*}}, \nonumber \\
\textrm{(ingoing waves)} ~~~~~~~~~~~~ \Psi_{4}^{(1)} & \sim &  \f{e^{- \imath \omega r^{*}}}{r^{*5}}.
\Ee
This results coincided with the peeling theorem \cite{11} that the gravitational perturbation $\Psi_{4}$, proportional to $1/r$ only, falls off slowly enough to be non-zero near the infinity.
Interestingly, the potential has both real and imaginary parts.
\be
V^{\textrm{NP}}(x)=\f{4\imath \omega r^{*}}{(r^{*2} +b_{0}^{2})}+\f{(l-1)(l+2)}{(r^{*2} +b_{0}^{2} )} + \f{6 b_{0}^{2}}{(r^{*2} +b_{0}^{2} )^{2}}.
\ee
We demonstrated the potential $V_{l}^{\textrm{NP}}$ in Fig.~1. As clearly shown in the final perturbed equation, the gravitational perturbation can be expressed in terms of one dimensional Schr\"{o}dinger equation with potential barrier. We considered the real part and the imaginary part separately because we should look into the meaning of the imaginary part of the potential in Schr\"{o}dinger equation further. The values of the real part of the effective potential are positive in every value of $l$ so that we concluded that the stability of the spacetime exists.

\begin{figure*}
\includegraphics[height=6cm]{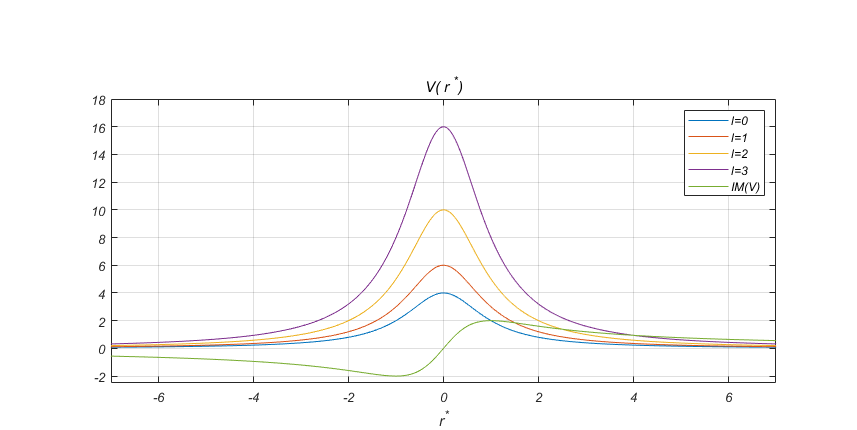}
\caption{\label{fig_1} A plot of the potentials of the gravitational perturbation, in terms of $r^*$ for $l=0, 1, 2$, and $3$. Plot the imaginary part separately. Here we set $b_0 =1$ and $\omega =1$.}
\end{figure*}
In our previous study \cite{12}, we derived the equation of the gravitational perturbation of Morris-Thorne wormhole spacetime in the form of the one dimensional Schr\"{o}dinger equation with potential barrier by using the linearized theory of gravity. In this study, we computed that by using Newman-Penrose formalism. In Fig.~2, we demonstrated the effective potentials $V_{l}^{\textrm{NP}}$, $V_{l}^{(-)}$ and $V_{l}^{(+)}$. $V_{l}^{(-)}$ represents the axial perturbation and $V_{l}^{(+)}$ demonstrates the polar perturbation, respectively.\\
In the case of the axial and the polar perturbation of the Morris-Thorne wormhole spacetime, the potential, $V_{l}^{(-)}$ and $V_{l}^{(+)}$, are real with no $\omega$ dependence so that the eigenvalue problem for $\omega^2$ is linear and self-adjoint.  However, the potentials for the axial perturbation are negative in some value of $l$ that indicates the instabilities of the spacetime. We found the interesting aspect in \cite{12} that the potentials of the axial perturbation became positive after we considered the interaction between the external perturbation and the exotic matter, constituting the Morris-Thorne wormhole. It was inevitable because the wormhole should be filled with the exotic matter, not vacuum \cite{13}. We demonstrated the results in Fig.~3. The potential $V^{\textrm{ex}}$ indicated the potential the interaction between the external perturbation and the background exotic matter.\\

\begin{figure*}
\includegraphics[height=6cm]{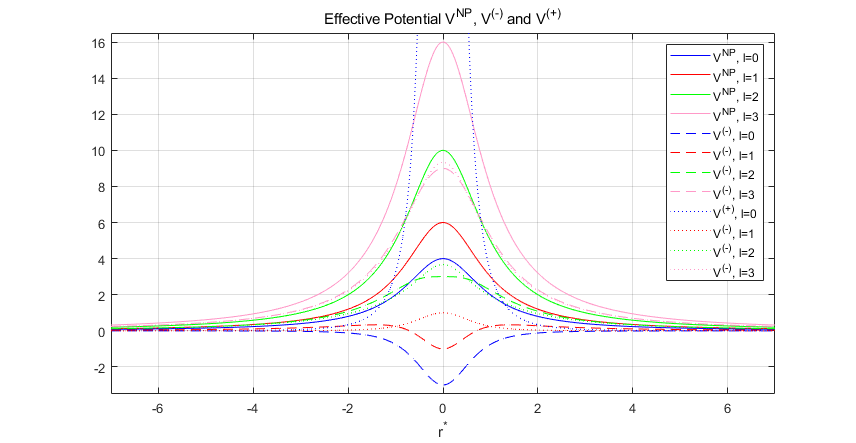}
\caption{\label{fig_2} A plot of the potentials of the gravitational perturbation, $V_{l}^{\textrm{NP}}$, $V_{l}^{(-)}$ and $V_{l}^{(+)}$, in terms of $r^*$ for $l=0, 1, 2$, and $3$. Here we set $b_0 =1$ and $\omega =1$.}
\end{figure*}

\begin{figure*}
\includegraphics[height=6cm]{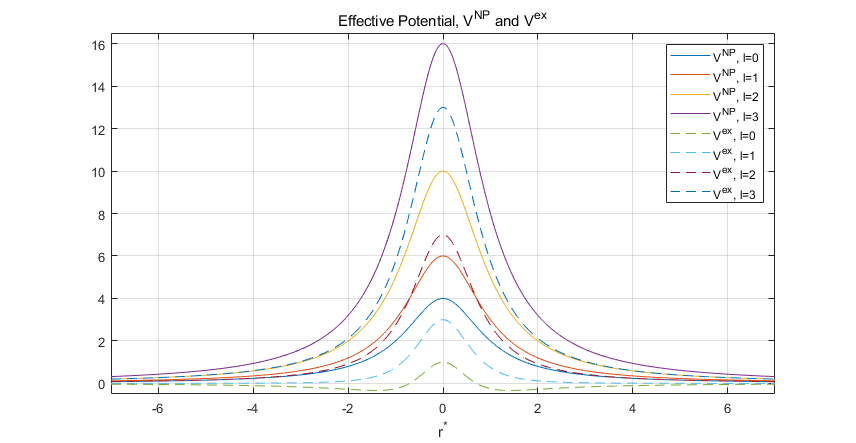}
\caption{\label{fig_3} A plot of the potentials of the gravitational perturbation, $V_{l}^{\textrm{NP}}$ and $V_{l}^{ex}$, in terms of $r^*$ for $l=0, 1, 2$, and $3$. Here we set $b_0 =1$ and $\omega =1$.}
\end{figure*}

\section{Perturbation and Radially In-falling Pressureless Dust Matter}
The advantage of adapting the Newman-Penrose Formalism to gravitational perturbation equation is to manipulate a source term that perturbs the given wormhole spacetime. In this paper, we assumed that the source is the pressureless dust matter.
The stress-energy tensor $T_{\mu \nu}$ of the dust is given as
\be
T_{\mu \nu} = \rho u_{\mu} u_{\nu}
\ee
where $\rho$ is the rest mass density profile and $u_{\mu}$ is the four velocity of the matter. The four velocity is a vector tangent to the world line of the particle, defined as
\be
u^{\mu} = \vec{u} \cdot e^{\mu} =\f{dx^{\mu}}{d \tau}
\ee
where $\tau$ is the proper time.\\
We assumed that the dust is radially in-falling into the wormhole throat so that the four velocity of the dust has only temporal and radial components
\be
u^{\mu} = \left( u^{t} , u^{r} , 0 , 0 \right)
\ee
which are functions of $r$ and $t$ only.\\
Therefore, the non-vanishing component of the energy stress tensor is $T_{kk}$, the projection of the energy stress tensor along the light cone in the $k$-direction
\Be
T_{km^{*}} &=& T_{m^{*} m^{*}} = 0 \nonumber \\
T_{kk} &=& \rho u_{k} u_{k}
\Ee
We projected on to the tetrad frame and obtained
\be
T_{kk} = \left( u_{t} -\sqrt{1-\f{b_{0}^{2}}{r^{2}}} u_{r} \right)^{2} \rho.
\ee
As a result, the source term $T_{4}$ is given by
\be
T_{4} = -\f{1}{2r^{2}} \left( u_{t} -\sqrt{1- \f{b_{0}^{2}}{r^{2}} } u_{r} \right)^{2} {\bar{\partial}}_{-1} {\bar{\partial}}_{0} \rho .
\ee
We assumed that the density profile was decomposed into the temporal-radial part and the angular part, in terms of the usual spherical harmonics with zero spin-weight
\be
\rho = \sum_{l,m} {\rho_{lm} (t,r) Y_{0}^{lm} (\theta, \phi)}.
\ee
The four velocity of the radially in-falling pressureless dust matter were given by
\Be
u^{t} &=& E,\nonumber \\
u^{r} &=& \pm \sqrt{ \left(1- \f{b_{0}^{2}}{r^{2}} \right) \left(u^{t^2} -1 \right) }.
\Ee
We took the minus sign for the $u^{r}$ because we assumed that the matter was only in-falling.\\
Using the four velocity, we could not only determine the velocity of each particle(dust) of fluid at a given time and position, but also compute the coordinate velocity, $v^{r}$, defined as
\be
\f{u^{r}}{u^{t}} = \f{\partial x }{\partial \tau} \f{\partial \tau}{\partial t} = \f{\partial x}{\partial t} = v^{r}.
\ee
We computed the continuity equation for the current vector $J^{\mu} = \rho u^{\mu}$ to describe the evolution of the fluid
\Be
J_{~;\mu}^{\mu} = J_{~,\mu}^{\mu} + \Gamma_{\lambda \mu}^{\mu} J^{\lambda} =0,\nonumber \\
\left[ \f{\partial }{\partial t} + v^{r} \left( \f{\partial }{\partial r} + \f{2}{r} \right) \right] \rho_{lm} (r,t) = 0.
\Ee
The above continuity equation is irregular at $r=0$ so that we conducted the coordinate transformation using proper radial distance $r^{*}$ and rewrote the equation
\be
\f{\partial \rho_{lm}}{\partial t} -\sqrt{\f{u^{t^2} -1}{u^{t^2}}} \f{\partial \rho_{lm}}{\partial r^{*}} - \f{2 r^{*}}{(r^{*2} +b_{0}^{2})} \sqrt{u^{t^2} -1} \rho_{lm} =0.
\ee
We finally substituted the source term, $T_{4}$, into the mater perturbed equation we obtained in the previous section and obtain the following.
\Be
&&\square_{tr}^{\Phi_{4}} R_{lm} - (l-1)(l+2) R_{lm} + K \sqrt{r^{*2} +b_{0}^{2}} \left( \sqrt{u^{t^2} -1} -u^{t} \right)^{2} \times \nonumber \\
&& \hskip 4cm \sqrt{(l-1)l(l+1)(l+2)} \rho_{lm} =0.
\Ee
As shown in the above equation, the gravitational perturbation, induced by the radially in-falling matter should be quadrupole $(l=2)$ at least.\\
By solving this equation numerically, varying the energy profile under the different assumptions, we may become one step closer to understand the nature of gravitational perturbation in wormhole spacetime.
\subsection{A Single Pulse of Dust Matter}
We assumed that the energy density profile of the pressureless dust matter consisted of two spherical harmonic modes,
\be
\rho (r,t) = \rho_{00} (r,t) Y_{0}^{00} + \rho_{lm} (r,t) Y_{0}^{lm},
\ee
and that the matter radially fell into the wormhole throat as a form of a single Gaussian pulse,
\be
\rho_{lm} (r^{*} ,0)=A_{0} e^{-(r^{*} -r_{0})^{2}/\sigma^{2}},
\ee
where $A_{0}$ is the initial amplitude of the Gaussian pulse, $r_{0}$ is the initial position of the pulse and $\sigma$ is the width.
The evolution of the energy density profile for a single Gaussian pulse is demonstrated in the Fig.~4. We considered the quadrupole mode, $l=2$, and assumed that the initial amplitude $A_{0} = 10^{-5}$, the width of pulse $\sigma=0.5$ and the energy $u^{t} = E = 1.1$.\\
\begin{figure*}
\includegraphics[height=6cm]{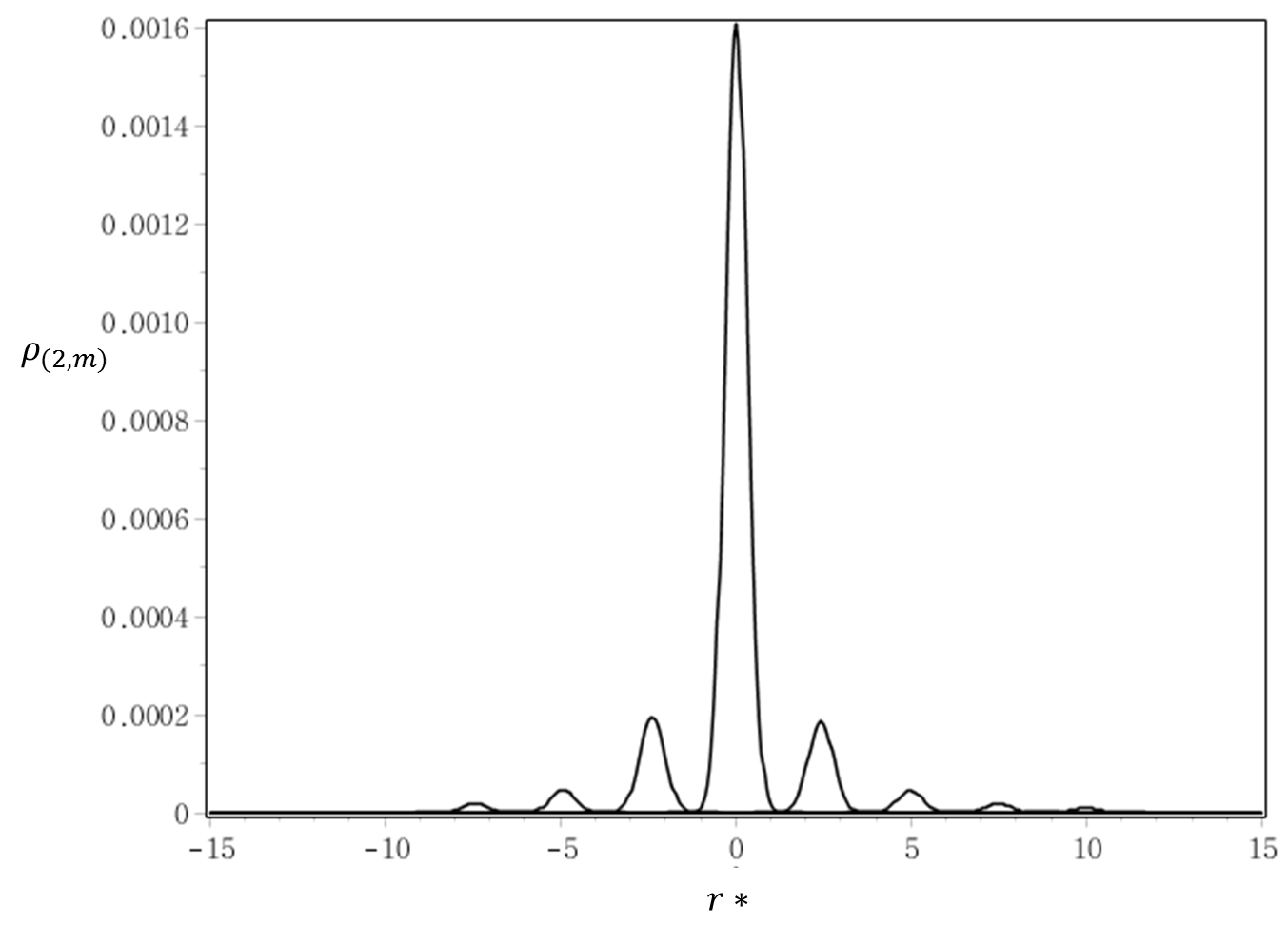}
\caption{\label{fig_4} The evolved density profile for a single pulse, located at $r_{0}=10$ initially, plotted every $t = 6\textrm{s}.$ The Gaussian has an initial amplitude $A_{0} = 10^{-5}$, an initial width $\sigma=0.5$.}
\end{figure*}
We substituted the density profile to the radial perturbation equation and solved it numerically using the computer algebra system Maple 18. We plotted the results in the Figs.~5 and ~6.

\begin{figure*}
\includegraphics[height=6cm]{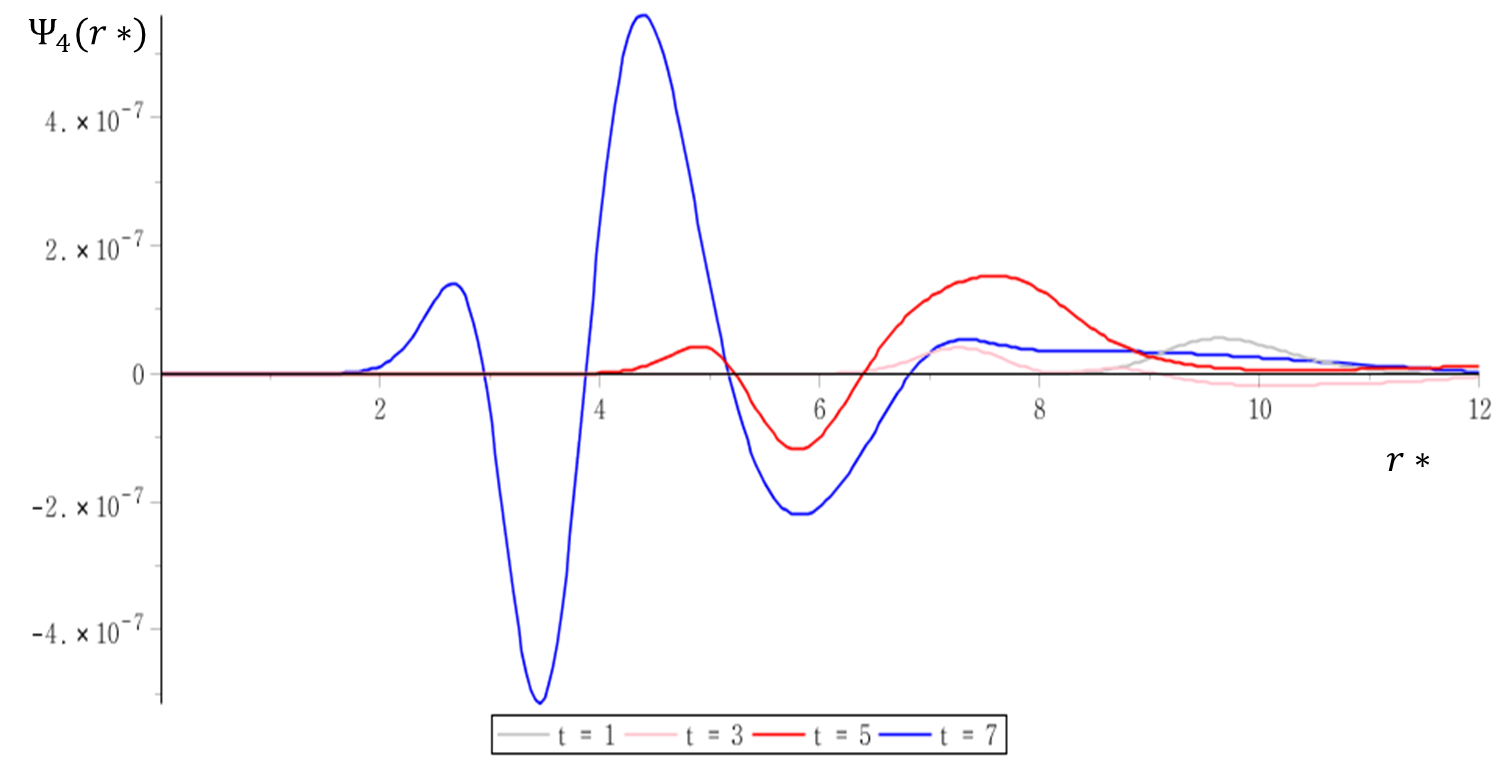}
\caption{\label{fig_5} The gravitational perturbation $\Psi_{4}^{(1)}$, induced by the single dust matter with $\rho_{20}$, propagating inward. }
\end{figure*}

\begin{figure*}
\includegraphics[height=6cm]{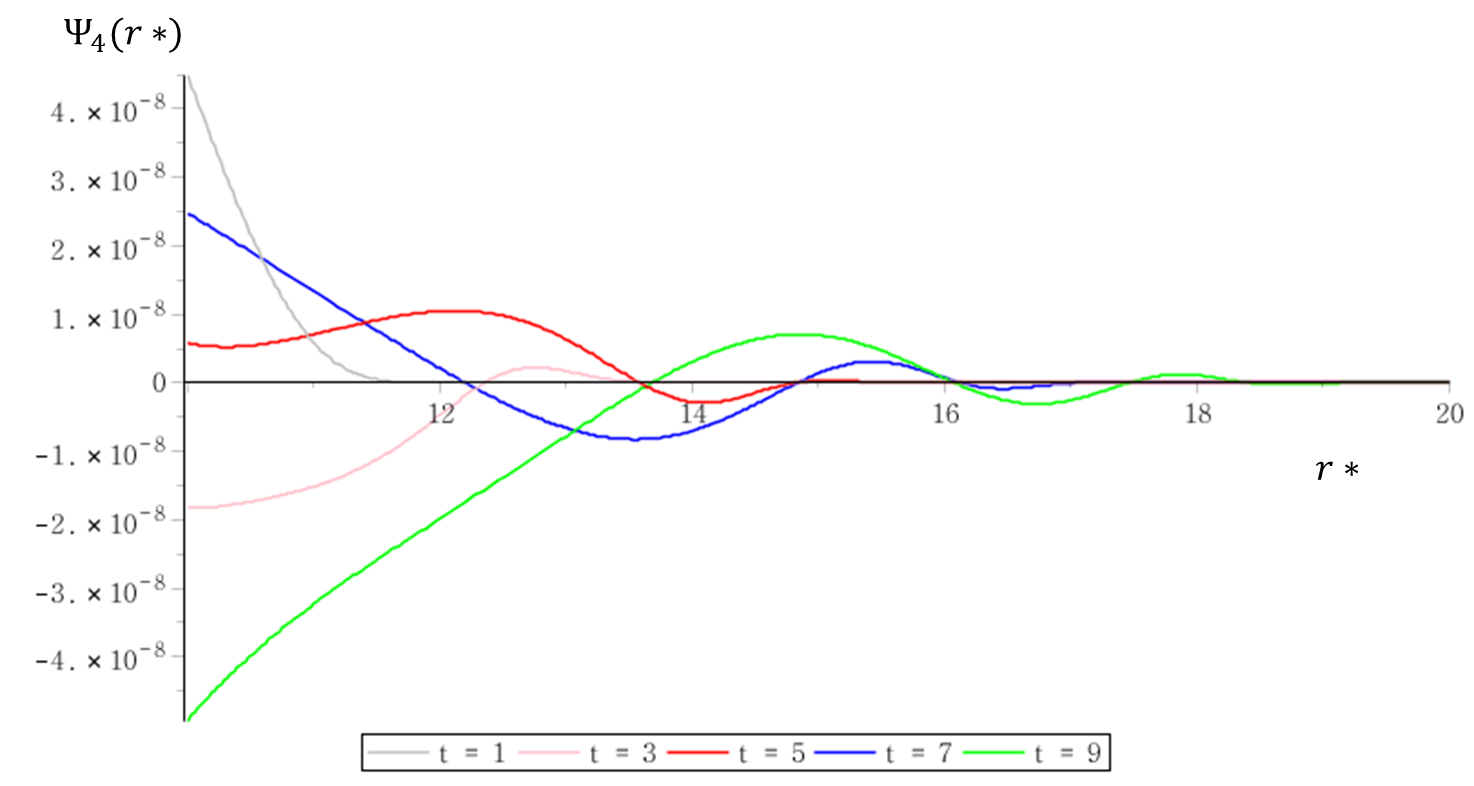}
\caption{\label{fig_6} The gravitational perturbation $\Psi_{4}^{(1)}$, induced by the single dust matter with $\rho_{20}$, propagating ourward. }
\end{figure*}

We set the initial condition as
\be
R_{02} (r^{*} ,0)=0, ~~~~~~~~~ \f{\partial R_{02}}{\partial t} |_{t=0} =0
\ee
assuming there was no perturbation at the initial time $(t=0)$.\\
As a single pulse of dust matter evolved toward the throat of the wormhole, located at $r^{*} =0$, it perturbed the background spacetime and yielded the gravitational perturbation. The amplitude of the gravitational perturbation grew faster not only because the curvature of the wormhole spacetime became steeper as approaching to the throat, but also because the time dependence was proportional to $e^{\imath \omega t}$. What we were interested in was the outgoing gravitational perturbation away from the throat, as shown in Figure ~6. The perturbed radial function $R_{20}$ propagated outward as time passes while the amplitude of the gravitational perturbation decreased, corresponding to the damped oscillation.

\section{Conclusion}
In this paper, we investigate the gravitational perturbation of the Morris-Thorne wormhole by using the Newman-Penrose formalism. Before applying the tetrad formalism, we had needed to check the characteristics of the Morris-Thorne wormhole spacetime whether it was suitable and confirmed it positively because the spacetime is a type D spacetime in Petrov classification. We found the null tetrad basis for the spacetime and computed $\Psi_{4}$ in the first (linear) order, corresponding to the outgoing gravitational perturbation. The radial equation contained the source term so that we could easily manipulate it which caused the gravitational perturbation. In other words, this approach provided us with a more effective and intrinsic way to manipulate the external source that produces the gravitational perturbation in wormhole spacetime.\\
We first considered the case with no external disturbance $(T_{4} = 0)$, anticipating that the result we would obtain might correspond to our previous work \cite{12}, the gravitational perturbation obtained by using the Linearized theory of gravity. However, an interesting and important difference in the imaginary part of the effective potential appeared in the radial perturbed equation. We suspected that the complex term appeared because of the characteristics of the coordinate bases since some null vectors of tetrad basis are defined as complex.\\
We then extended our study to derive the perturbed equation, including the source term, a single Gaussian pulse of pressureless dust matter and solved it numerically. Even though the problem should be solved numerically, we could obtain the gravitational perturbation produced by a single pulse, radially falling into Morris-Thorne wormhole by using Newman-Penrose formalism in this paper.

\begin{acknowledgments}
This work was supported by National Research Foundation of Korea (NRF) funded by the Ministry of Education (2017-R1D1A1B03031081).

\end{acknowledgments}

\end{document}